\documentclass[10pt,journal,final,finalsubmission,twocolumn,amsart]{IEEEtran}

\usepackage{amsmath}
\usepackage{amstext}
\usepackage{amssymb}
\usepackage{latexsym}
\usepackage{times}
\usepackage{array}
\usepackage{setspace}
\usepackage{nicefrac}
\usepackage{booktabs}
\usepackage{bbold}
\usepackage{algorithm,algpseudocode}
\usepackage{lipsum}
\usepackage{ifpdf}
\usepackage{mathtools}
\usepackage{cuted}
\usepackage{multicol}
\usepackage{nicefrac}
\usepackage[utf8]{inputenc}
\usepackage[english]{babel}
\usepackage{amsthm}
\usepackage{hhline}
\usepackage{boldline}
\usepackage{placeins}
\addto\captionsenglish{}

\newtheorem{thm}{Claim}

\newcommand{\ith}[1]    {{#1}^{\underline{ \text{th}}}}
\newcommand{\paren}[1]{\left({#1}\right)}

\newcommand{\braces}[1]{{\left\{ {#1}\right\}}}
\newcolumntype{L}{>{\centering\arraybackslash}m{3cm}}

\ifpdf
\usepackage[pdftex]{graphicx} 
\usepackage{epstopdf}         
\else
\usepackage{graphicx}         
\fi

\ifpdf
\DeclareGraphicsExtensions{.pdf, .jpg, .tif}
\else
\DeclareGraphicsExtensions{.eps, .jpg}
\fi

\usepackage{rotating}

\title{Real Signal Equalization for OQAM}

\author{Md Navid Akbar, and Mohammad Saquib 
\thanks{Md N. Akbar and M. Saquib are with the Department of Electrical Engineering, The University of Texas at Dallas, 800 W. Campbell Road, Richardson, Texas 75080-3021. Emails: mdnavid.akbar@utdallas.edu, saquib@utdallas.edu} \addtocounter{footnote}{-1}\footnotemark
}

\begin{document}

\maketitle
\vspace{-2mm}
\begin{abstract}
This correspondence proposes the use of a real-only equalizer
(ROE), which acts on real signals derived from the received offset quadrature amplitude modulation (OQAM) symbols. For the same fading channel, we prove that
both ROE and the widely linear equalizer (WLE)
yield equivalent outputs. Hence, these exhibit the same performance. Our complexity analysis finds that depending on the frame length,
ROE can be computationally less complex, and save significant
signal processing time over WLE. In the adaptive normalized least mean square implementation, ROE performs better with lower complexity than its counterpart, for a given number of pilot bits.
\end{abstract}
\vspace{-1mm}

\begin{IEEEkeywords}
Real-only equalizer (ROE), widely linear equalizer (WLE), adaptive equalization, computational complexity.
\end{IEEEkeywords}

\vspace{-2mm}
\section{Introduction}
\vspace{-2mm}
Offset quadrature amplitude modulation (OQAM) schemes are widely used in the wireless communication scene, such as in LTE \cite{OQAM-one}, as well as in 5G protocols \cite{OQAM-two}. A particular case of the OQAM scheme is the offset quadrature phase-shift keying (OQPSK) signal. In applications where the transmitter power is limited, OQPSK is used, as it is more compatible than linear modulations, with RF power amplifiers operating in full saturation \cite{rice}.
Examples of systems employing OQPSK include
wideband code division multiple access (WCDMA) mobile systems \cite{mirbagheri},
satellite communication \cite{satellite}, and the IEEE 802.15.4 (Zigbee) networking protocol \cite{zigbee}.

In wireless communication, a signal undergoes multipath effects.
The most common form of multipath mitigation is equalization, performed at the receiver end.
Equalization for OQAM has been studied extensively, such as in \cite{bello-pahlavan}.
Picinbono and Chevalier in \cite{picinbono} first theorized that for complex signals,
widely linear equalizer (WLE) provides a better estimate compared to the classical minimum mean square error (MMSE) linear equalizer (LE), when the pseudo-autocorrelation of the received symbols is non-zero.
WLE has subsequently become the most recommended receiver for the OQPSK scheme \cite{lampe}.

WLE performs two linear transformations: one on the sampled matched filter output, and the other on its conjugate.
It operates by inverting the complex channel autocorrelation matrix.
Depending on the size, this matrix inversion could be computationally quite intense.
This is where the proposed real-only equalizer (ROE) could offer an advantage.
A pre-processing multiplier inside ROE converts the sampled matched filter output to a real-valued input for ROE.
Next, ROE inverts the resulting (real-valued) channel autocorrelation matrix. This fact assists ROE to be a reduced complexity equalizer, and forms the basis for the following key results: 1) ROE and WLE are proven to provide equivalent outputs. Hence, they will exhibit the same bit error rate (BER) performance. 2) Presented analysis implies that ROE is expected to be computationally less intense. For practical frame lengths, numerical results demonstrate that ROE takes significantly less computational time, compared to WLE. 3) In adaptive equalization involving the normalized least mean square (NLMS) algorithm, ROE is found superior to WLE, in terms of both complexity and BER performance (for a given number of training bits).
4) Performance of ROE is analyzed and found to outperform the classical LE in terms of average MMSE.

The notations used in this paper are as: a boldface lowercase variable (e.g., $\mathbf{v}$) denotes a column vector,
a boldface uppercase variable (e.g., $\mathbf{M}$) denotes a matrix;
the transpose and Hermitian operations are $\mathbf{v}^\top$ ($\mathbf{M}^\top$) and $\mathbf{v}^\dag$ ($\mathbf{M}^\dag$), respectively;
$\mathbf{v}^\ast$ and $\mathbf{M}^\ast$ denote complex-conjugates; $E[\cdot]$ is the expectation operator; $\text{diag}[\dots]$ is a diagonal matrix containing the elements specified; $\text{tr}\{\mathbf{M}\}$ contains the trace of $\mathbf{M}$; $\textbf{0}_n$ is a zero vector of length $n$; $\mathbf{v}[a:b]$ contains elements $a$ through $b$ in $\mathbf{v}$;
$\|\mathbf{v}\|_2$ is the $\ell_2$ norm of $\mathbf{v}$;
for a square matrix, $\mathbf{M}^{-1}$ denotes the matrix inverse; $\mathbf{I}$ is the identity matrix whose dimensions are determined by context.
\vspace{-2mm}
\section{System Model}
\vspace{-2mm}
An OQAM signal $x(t)$ consisting of $2N$ symbols (corresponding to $M$-QAM symbols, where $M=N^2$ and $N=2,4,8,...$) is transmitted through a channel with impulse response $h_c(t)$, following a pulse shaping and modulating filter $p(t)$. In practice, this filter is a unit-energy, square-root Nyquist pulse. For simplicity of analysis and simulation, a rectangular window-based filter is considered in this work.
At the output of filter $p(t)$, we have the continuous signal $x(t)$. A total of $2\nu$ symbols are transmitted.
This  $x(t)$ travels through the channel and gets accompanied by an additive noise $z(t)$.
After application of a low-pass anti-aliasing filter with impulse response $h_a(t)$ and bandwidth $1/(2T_s)$ at the receiver, as described in \cite{riceEQ}, an analog-to-digital (A/D) converter produces $T_s$ symbol-duration spaced samples
\vspace{-1mm}
\begin{align}\label{r-n}
  r(n) &= \sum_{k=0}^{L_s} h(k)x(n-k) + z(n),
\end{align}
where
$h(n) = \left. h_c(t) \ast h_a(t) \right|_{t = nT_s}$,
$x(n) = x(nT_s)$ and $\quad z(n) = \left. z(t) \ast h_a(t) \right|_{t = nT_s}$.

The constellation points for a $2N$-OQAM modulation are
\vspace{-1mm}
  \[
    x(n)=\begin{cases}
                  a(n) \qquad \;\, n \; \text{is even}\\
                  ja(n) \qquad n \; \text{is odd},
        \end{cases}
  \]

where for $A>0$,  $a(n) \in \braces{\pm A, \pm 3A, \pm\paren{N-1} A}$ is the amplitude of $\ith{n}$  symbol,
and the causal discrete-time channel has support on $0 \leq n \leq L_s$ (i.e., the discrete-time equivalent channel has $L_s + 1$ taps, each spaced by $T_s$).
In addition,  $a(n)$ is an equally likely independent symbol sequence.
The additive noise $z(t)$ is modeled as a zero-mean, circularly symmetric complex-valued white Gaussian random process whose real and imaginary parts have the same power spectral densities $N_0$ W/Hz.
We then apply a filter $h^\ast(-n)$, matched to the discrete-time equivalent channel, to produce:
\begin{align}
	y(n) &= x(n)\ast \underbrace{h(n)\ast h^\ast(-n)}_{\gamma(n)} +
			\underbrace{z(n)\ast h^\ast(-n)}_{v(n)}.
\end{align}
\vspace{-1mm}
\begin{align}
\therefore y(n)= \sum_{k=-L_s}^{L_s} \gamma(k)x(n-k) + v(n),
\end{align}
where
$v(n)$ is a complex-valued correlated noise sequence, and $x(n) = 0$ for $n < 0$ or $n >= 2\nu$.
The vector formed from stacking $y(0), \ldots , y((2\nu-1)T_s)$ becomes
\begin{equation}\label{y-disc}
  \textbf{y} = \boldsymbol{\Gamma} \textbf{x}+\textbf{v},
\end{equation}
where $\boldsymbol{\Gamma}$ is the $2\nu \times 2\nu$ complex channel autocorrelation matrix formed from the complex-valued $\gamma (k)$ as
\[
\boldsymbol{\Gamma} =
\begin{bmatrix}
{\gamma}(0)   &{\gamma}(-1)   &\dots           &{\gamma}(-L_b)  &\dots           &0\\
{\gamma}(+1)  & {\gamma}(0)   &\dots           &\dots             &\dots           &0\\
\vdots        &  \vdots       &  \ddots        &\vdots            &\ddots          &\vdots\\
0             &  0            &\dots           &\dots             &\dots           &{\gamma}(0)
\end{bmatrix},
\]
$\textbf{x}$ and $\textbf{v}$ are $2\nu \times 1$ complex vectors formed from $x(n)$ and $v(n)$, respectively. The autocorrelation matrix of the noise vector $\textbf{v}$ is $2N_0 \boldsymbol{\Gamma}$ and $E(\textbf{v})=0$.
\vspace{-2mm}
\section{Real-Only Equalizer}
\vspace{-1mm}
\subsection{Receiver Structure and Equivalence}
\vspace{-1mm}
At the $\ith{n}$ matched filter output, when $n$ is even, the imaginary part carries no information regarding the desired symbol $a(n)$, and similarly $n$ is odd, the real part does not have any information of $a(n)$ \cite{rice}.
Therefore, ROE eliminates the non-useful parts from the outputs of the complex matched filter, while estimating the transmitted symbols by processing the remaining useful parts.
Consequently, we are interested in the real-valued vector $\textbf{y}_a$ derived from the useful parts of $\textbf{y}$ as
\begin{equation}\label{ya-stack}
\textbf{y}_a = \big[\text{Re}\{y(0)\}\quad \text{Im}\{y(1)\} \dots
 \, \text{Im}\{y(2\nu-1)\} \big]^\top.
\end{equation}
To derive the upcoming results, we first define two matrices:
\begin{align*}
  \textbf{A} &= \text{diag}\left[ \frac{1}{2},\frac{-j}{2},\frac{1}{2}, \dots , \frac{-j}{2} \right]; & \textbf{B} &= \text{diag}\left[ 1,-1,1, \dots,-1 \right];
\end{align*}
which have the following properties:
\begin{align*}
  \textbf{A}^{\dagger} =\textbf{A}\textbf{B}; && \textbf{A}^{-1} =4\textbf{A}\textbf{B}; && \textbf{A} = \textbf{B}\textbf{A}\textbf{B}; && \textbf{B}=\textbf{B}^{-1}; && \textbf{B}^2 = \textbf{I}.\notag
\end{align*}
Equation \eqref{ya-stack} can be rewritten using the relation between OQAM symbols and binary bits, $\textbf{x}=2\textbf{A}^{\dagger}\textbf{a}$, and (\ref{y-disc}) as
\begin{equation}
  \textbf{y}_a = \textbf{A}(\textbf{y}+\textbf{B}\textbf{y}^*) = \boldsymbol{\Gamma}_a \textbf{a}+\textbf{v}_a, \label{ROE-y-alt}
\end{equation}
where $\textbf{a}$ is the $2\nu \times 1$ real vector formed from $a(n)$; $\textbf{v}_a$ is the equivalent, real-valued correlated noise with zero mean, autocorrelation $2N_0 \boldsymbol{\Gamma}_a$; and $\boldsymbol{\Gamma}_a$ is the equivalent, real-valued channel autocorrelation matrix given by
\begin{equation}\label{Gamma-a}
    \boldsymbol{\Gamma}_a = 2\textbf{A}(\boldsymbol{\Gamma}+\textbf{B}\boldsymbol{\Gamma}^\top\textbf{B}) \textbf{A}^{\dagger} = \hat{\boldsymbol{\Gamma}} + \bar{\boldsymbol{\Gamma}};
\end{equation}
where
\vspace{-1mm}
$$
\hat{\boldsymbol{\Gamma}} = 2(\textbf{A})\times \frac{1}{2}\boldsymbol{\Gamma}\times 2(\textbf{A})^{\dagger}; \: \:
\bar{\boldsymbol{\Gamma}} = 2(\textbf{AB})\times \frac{1}{2}\boldsymbol{\Gamma}^\top \times 2(\textbf{AB})^{\dagger}.
$$
\vspace{-2mm}

Let us focus on the equalizers next. First, we take a look into the classical LE. It provides an estimate ($\hat{\textbf{x}}_\text{LE}$) of any received OQAM symbol, under the MMSE criterion, as
\begin{equation}\label{Filter-LE}
  \hat{\textbf{x}}_\text{LE}  \: \:=  \: \: \left[\boldsymbol{\Gamma} + \frac{\sigma^2}{A_N^2} \textbf{I}\right]^{-1}\textbf{y}  \: \:= \: \: \left[\boldsymbol{\Gamma} + \sigma^2 \textbf{I}\right]^{-1}\textbf{y},
\end{equation}
where $A_N^2=\frac{2A^2}{N}\braces{1^2+3^2+\ldots+(N-1)^2} $, is the average symbol energy (which, for convenience, set to 1) and  $\sigma^2 = 2N_0$. The signal to noise ratio (SNR) thus becomes $1/ \sigma^2$.
For ROE, noise power will be $\sigma^2 /2$ for the noise without imaginary parts, the channel autocorrelation matrix becomes $\boldsymbol{\Gamma}_a$, and the effective new input is $\textbf{y}_a$. The output of ROE ($\hat{\textbf{x}}_\text{ROE}$) can thus be written as that of LE in (\ref{Filter-LE})
\begin{align}\label{estimate-a}
  \hat{\textbf{x}}_\text{ROE} &= \hat{\textbf{a}} = \left[\boldsymbol{\Gamma}_a + \frac{\sigma^2}{2} \textbf{I}\right]^{-1}\textbf{y}_a.
\end{align}
Next, we will establish the relation between the outputs of ROE and WLE, under known channel state information (CSI).
\begin{thm}\label{thm-1}
The output of ROE is the real-only equivalent of the output of WLE ($\hat{\mathbf{x}}_\text{WLE}$), and is given by
$\hat{\mathbf{x}}_\text{ROE} = 2\mathbf{A}\times\hat{\mathbf{x}}_\text{WLE}$.
\end{thm}
\noindent
\textit{Proof.} Using (\ref{ROE-y-alt}) and (\ref{Gamma-a}), $\hat{\textbf{x}}_\text{ROE}$ in (\ref{estimate-a}) may be rewritten as
\begin{align}\label{ROE-imp1}
  \hat{\textbf{x}}_\text{ROE}  &=\left[2\textbf{A}(\boldsymbol{\Gamma} + \textbf{B} \boldsymbol{\Gamma}^\top \textbf{B})\textbf{A}^{\dagger} + \frac{\sigma^2}{2} \textbf{I}\right]^{-1} [\textbf{A}\textbf{y}+\textbf{A}\textbf{B}\textbf{y}^*].
\end{align}
\noindent
Applying the properties of $\textbf{A}$ and $\textbf{B}$ discussed earlier, we get
\begin{align}
    \hat{\textbf{x}}_\text{ROE}  &=\left[ \frac{1}{2} \left\{\boldsymbol{\Gamma}+ \textbf{B}\boldsymbol{\Gamma}^\top\textbf{B}+\sigma^2 \textbf{I}\right\}   (4\textbf{A}\textbf{B})\right]^{-1}(\textbf{y}+\textbf{B}\textbf{y}^*) \notag\\
    &=2\textbf{A}(\boldsymbol{\Gamma}+\textbf{B}\boldsymbol{\Gamma}^\top \textbf{B}+\sigma^2 \textbf{I})^{-1}(\textbf{y}+\textbf{B}\textbf{y}^*).\label{theo-1-proof}
\end{align}
Structure of the WLE is derived in the Appendix. From there, replacing $\textbf{C}$ and $\textbf{D}$ in (\ref{theo-1-proof}) completes the proof:
$$ \hat{\textbf{x}}_\text{ROE} =2\textbf{A}[\textbf{C}\textbf{y} + \textbf{D}\textbf{y}^*]=2\textbf{A}\times\hat{\textbf{x}}_\text{WLE}.\: \: \: \: \: \: \: \: \: \: \:  \square $$
The result of Claim 1 will also hold when ROE and WLE will be implemented using estimated CSI.
Inter-symbol interference (ISI)
seen by the symbols located at the beginning or ending of a frame is very different than
that seen by the symbols in the middle of the frame. It is due to the finite frame length. ISI is the main contributing
factor behind bit error, and it is different for the edge symbols than the symbols close to the
center.
When the transmission block length is sufficiently large, MMSE of every single symbol can be reasonably approximated by average MMSE.

In the next section, we will analyze the performance of ROE using average MMSE as the performance metric. Even though BER is a more desirable metric, the closed-form expression of BER is quite difficult to analyze for MMSE equalizers in multipath channels.
Nonetheless, the equalizer that yields a lower MMSE is expected to perform better in terms of BER \cite{BER-MSE}.
\vspace{-3mm}
\subsection{Average MMSE Analysis of ROE}
\vspace{-2mm}
The instantaneous error vector for ROE is given by
\vspace{-1mm}
\begin{align}
\textbf{e} &= (\textbf{a}-\hat{\textbf{a}})
 = \textbf{a} - \left[\boldsymbol{\Gamma}_a+\frac{\sigma^2}{2}\textbf{I}\right]^{-1} [\boldsymbol{\Gamma}_a\textbf{a}+\textbf{v}_a]\notag \\
& = \Bigg[\textbf{I} - \left(\boldsymbol{\Gamma}_a+\frac{\sigma^2}{2} \textbf{I}\right)^{-1}\boldsymbol{\Gamma}_a\Bigg]\textbf{a} -\left(\boldsymbol{\Gamma}_a+\frac{\sigma^2}{2} \textbf{I}\right)^{-1}\textbf{v}_a,\label{error-vector}
\end{align}
and its corresponding error covariance matrix is
\vspace{-1mm}
\begin{equation}\label{err-cov}
\boldsymbol{\Sigma} = E[\textbf{e}\textbf{e}^{\dagger}].
\end{equation}
\noindent
Now we use \eqref{err-cov} to express average MMSE of ROE $(\text{MMSE}_\text{av,ROE})$ over all $2\nu$ symbols as
\begin{equation}\label{trace-avg}
  \text{MMSE}_\text{av,ROE}=\frac{1}{2\nu} \text{tr} \left\{\boldsymbol{\Sigma}\right\}.
\end{equation}
Using \eqref{Gamma-a} and \eqref{error-vector} in  \eqref{err-cov}, and then applying to \eqref{trace-avg}, we get
\begin{align}\label{MMSE-RO}
  \text{MMSE}_\text{av,ROE} &=\frac{\sigma^2}{4\nu}\text{tr}\left\{\left[\hat{\boldsymbol{\Gamma}}+ \bar{\boldsymbol{\Gamma}}+\frac{\sigma^2}{2}\textbf{I}\right]^{-1}\right\}.
\end{align}
Similarly, the expression for average MMSE of LE ($\text{MMSE}_\text{av,LE}$) can be written as
\begin{equation}\label{MMSE-LE-Mat}
  \text{MMSE}_\text{av,LE}=\frac{\sigma^2}{2\nu}\text{tr}\left\{\left[\boldsymbol{\Gamma}+ \sigma^2\textbf{I}\right]^{-1}\right\}.
\end{equation}
The next claim establishes the relation between $\text{MMSE}_\text{av,ROE}$ and $\text{MMSE}_\text{av,LE}$.
\begin{thm}
$\text{MMSE}_\text{av,ROE}$ is less than, or equal to, $\text{MMSE}_\text{av,LE}$.
\end{thm}
\noindent
\textit{Proof.}
Let $\tilde{\boldsymbol{\Gamma}} =\hat{\boldsymbol{\Gamma}}+\frac{\sigma^2}{2}\textbf{I}$, and by using the matrix inversion lemma, we can rewrite (\ref{MMSE-RO}) as
\begin{align*}\label{MMSE-RO-simp}
\text{MMSE}_\text{av,ROE}&=\frac{\sigma^2}{4\nu}\text{tr}\{\tilde{\boldsymbol{\Gamma}}^{-1}- \tilde{\boldsymbol{\Gamma}}^{-1}(\bar{\boldsymbol{\Gamma}}^{-1}+ \tilde{\boldsymbol{\Gamma}}^{-1})^{-1}\tilde{\boldsymbol{\Gamma}}^{-1}\}.
\end{align*}
The autocorrelation matrix $\boldsymbol{\Gamma}$ is positive-definite, and thus diagonalizable. From \eqref{Gamma-a}, we realize that the diagonal elements of $\hat{\boldsymbol{\Gamma}}$ will be half of those in $\boldsymbol{\Gamma}$.
\noindent
Consequently, we can identify the first part of the above equation as
\begin{equation}\label{MMSE-LE}
  \text{MMSE}_\text{av,LE}=\frac{\sigma^2}{2\nu}\text{tr}\left\{\left[\boldsymbol{\Gamma}+ \sigma^2\textbf{I}\right]^{-1}\right\},
\end{equation}
\vspace{-1mm}
and denote the second part as
\begin{equation}\label{rho}
\rho=\frac{\sigma^2}{4\nu}\text{tr}\left[\tilde{\boldsymbol{\Gamma}}^{-1} \left(\bar{\boldsymbol{\Gamma}}^{-1} +\tilde{\boldsymbol{\Gamma}}^{-1}\right)^{-1}\tilde{\boldsymbol{\Gamma}}^{-1}\right]\geq 0,
\end{equation}
since both $\tilde{\boldsymbol{\Gamma}}$ and $\bar{\boldsymbol{\Gamma}}$ are also positive-definite matrices.
Substituting  (\ref{MMSE-LE}) and (\ref{rho}) back into the original equation involving the matrix inversion lemma completes the proof:
$$\text{MMSE}_\text{av,ROE} = \text{MMSE}_\text{av,LE}-\rho \leq \text{MMSE}_\text{av,LE}. \: \: \: \: \: \: \: \: \: \square$$
\noindent
This MMSE reduction (i.e. $\rho$) of ROE over LE can be zero:\\
\noindent
1) when the channel is perfectly noiseless, i.e. $\sigma^2$ is zero, and\\
2) in the case of real signal transmission over a real channel when $\boldsymbol{\Gamma}$ becomes $\boldsymbol{\Gamma}_a$ and noise power $\sigma^2$ changes to $\sigma^2/2$ in (\ref{MMSE-LE-Mat}). This result is expected. When all the system parameters are real, MMSE of WLE will also be the same as that of LE.
\vspace{-8mm}
\subsection{Alternative Implementation}
\vspace{-2mm}
Entire OQPSK received vector from \eqref{r-n} can be written as:
\begin{equation}
  \textbf{r} = 2\textbf{H}\textbf{A}^{\dagger} \textbf{a}+\textbf{z}=\hat{\textbf{H}}\textbf{a}+\textbf{z}, \label{r-ROE}
\end{equation}
where $\textbf{r}$ and $\textbf{z}$ are the $(2\nu+L_b) \times 1$ complex vectors formed from $r(n)$ and $z(n)$, respectively, and $\textbf{H}$ is the $(2\nu+L_b) \times 2\nu$ complex channel matrix obtained from appending discrete-time  channel vector $[\textbf{0}_n \; h(0)\cdots h(L_b) \; \textbf{0}_{(2\nu-n-L_b-1)}]^\top$ for the $\ith{n}$ transmission as columns.
The real and imaginary parts in (\ref{r-ROE}) can be separated as:
\begin{equation}\label{alt-ROE}
  \textbf{r}_{\text{R}} + j\textbf{r}_{\text{I}} = (\hat{\textbf{H}}_\text{R}+j\hat{\textbf{H}}_{\text{I}}) \textbf{a}+(\textbf{z}_\text{R}+j \textbf{z}_{\text{I}}),
\end{equation}
where $\hat{\textbf{H}}_\text{R}$ and $\hat{\textbf{H}}_\text{I}$ are real matrices, and $\textbf{z}_\text{R}$ and $\textbf{z}_\text{I}$ are real vectors.
Stacking them as an augmented vector and applying matched filter with real coefficients gives
\begin{gather}\label{step-1}
 \textbf{y}_b = \Big(\hat{\textbf{H}}_{\text{R}}^\top \hat{\textbf{H}}_{\text{R}}+ \hat{\textbf{H}}_{\text{I}}^\top \hat{\textbf{H}}_{\text{I}}\Big)\textbf{a}+ \Big(\hat{\textbf{H}}_{\text{R}}^\top \textbf{z}_{\text{R}} +\hat{\textbf{H}}_{\text{I}}^\top \textbf{z}_{\text{I}}\Big).
\end{gather}
Earlier in \eqref{ROE-y-alt}, ROE was realized by post-processing matched filter output $\textbf{y}$ to a real-valued stream.
By proving the following claim, we will demonstrate an alternative realization of ROE: by pre-processing the sampled received signal to a real-valued input, and then applying a matched filter as in \eqref{step-1}.
\begin{thm}\label{theo-3}
Linear transformation $[\boldsymbol{\Gamma}_a + \frac{\sigma^2}{2} \mathbf{I}]^{-1}$ on $\mathbf{y}_b$ from \eqref{step-1} will yield the same output,
as that obtained from $\mathbf{y}_a$ in \eqref{estimate-a}.
\end{thm}
\vspace{-1mm}
\noindent
\textit{Proof.} To prove the above claim, we only need to show that $\textbf{y}_a$ equals $\textbf{y}_b$.
Knowing that
$\hat{\textbf{H}}_{\text{R}}=(\textbf{H}\textbf{A}^{\dagger}+ \textbf{H}^*\textbf{A}^{\top}) \; \text{and} \;\hat{\textbf{H}}_{\text{I}}= j(\textbf{H}^*\textbf{A}^{\top}-\textbf{H}\textbf{A}^{\dagger}),$
\eqref{step-1} can be simplified to
\begin{align*}
\textbf{y}_b=2(\textbf{A}\textbf{H}^{\dagger} \textbf{H}\textbf{A}^{\dagger}+ \textbf{A}^*\textbf{H}^{\top} \textbf{H}^*\textbf{A}^{\top})\textbf{a}+ (\textbf{A}\textbf{H}^{\dagger}\textbf{n} +\textbf{A}^*\textbf{H}^{\top}\textbf{n}^*).
\end{align*}
Recognizing that $\textbf{A}^*=\textbf{A}\textbf{B}$, above equation can be written as
\begin{align}\label{step-3}
\textbf{y}_b=\textbf{A}[\boldsymbol{\Gamma}(2\textbf{A}^{\dagger}\textbf{a})+ \textbf{H}^{\dagger}\textbf{n}] +\textbf{A}\textbf{B}[\boldsymbol{\Gamma}^*(2\textbf{A}^{\top}\textbf{a})+ \textbf{H}^{\top}\textbf{n}^*].
\end{align}
Ultimately, we obtain our desired equivalence from \eqref{step-3} as
$$\textbf{y}_b=\textbf{A}[(\boldsymbol{\Gamma} \textbf{x}+ \textbf{v})+\textbf{B}(\boldsymbol{\Gamma} \textbf{x}+\textbf{v})^*]=\textbf{A}(\textbf{y}+\textbf{B}\textbf{y}^*)=\textbf{y}_a. \: \square $$
\vspace{-8mm}
\subsection{Adaptive Equalization}
\vspace{-1mm}
Guidelines for developing adaptive LE and WLE filters ($\textbf{w}_{\text{LE}}$ and $\textbf{w}_{\text{WLE}}$) are presented in \cite{haykin} and \cite{Adaptive_EQ_WLE}, respectively.
We develop an adaptive filter (based on NLMS) for ROE ($\textbf{w}_{\text{ROE}}$), using the result of pre-processing discussed in the previous section, and the proof of Claim \ref{theo-3}. 
Both NLMS and recursive least squares (RLS) are popular adaptive algorithms. 
In this work, we will primarily consider NLMS, which performs comparably to RLS but is computationally less complex \cite{adaptive-comp}.
An adaptive equalizer provides symbol-wise output, unlike the entire symbol vector decoding, described in the above sections.
The effective multipath channels are different for odd and even transmitted symbols, and hence we need two
adaptive algorithms to adapt ROE.
For procedural illustration, the ROE filter corresponding to even symbols will only be shown henceforward. Estimates for the $\ith{n}$ even symbol are given by the following relations:
\begin{gather}
  \hat{a}(n) = [\textbf{w}_{\text{ROE},n}]^\top
  \begin{bmatrix}
    \textbf{r}_{\text{R}}[n-mL_b:n+mL_b] \\
    \textbf{r}_{\text{I}}[n-mL_b:n+mL_b]
  \end{bmatrix},\notag
\end{gather}
\begin{align}
\therefore \hat{a}(n) ={\textbf{w}_{\text{ROE},n}}^\top \textbf{u}_{\text{ROE},n}.
\end{align}
Here length of each ROE filter and $\textbf{u}_{\text{ROE},n} \, (\ith{n} \,\, \text{input vector})$ is $(4m \times L_b+2) \times 1$, and filter length parameter $m \geq 3$ \cite{filter_len}. The $\ith{(n+2)}$ update of the ROE filter for the next even symbol becomes:
\vspace{-1mm}
\begin{align*}
  \textbf{w}_{\text{ROE},n+2} &= \textbf{w}_{\text{ROE},n}+\tilde{\mu}_\text{ROE}(n) \times \{a(n)-\hat{a}(n)\} \times \textbf{u}_{\text{ROE},n};
\end{align*}
where $\tilde{\mu}_\text{ROE}(n)= \mu_\text{ROE}/ (\delta_\text{ROE}+\|\textbf{u}_{\text{ROE},n}\|^2_2)$ is the time-varying step-size parameter, and $\mu_\text{ROE}$ and $\delta_\text{ROE}$ are constants of choice.
After all the equalizers (LE, WLE and ROE) have been trained using pilot symbols, they are used to extract information symbols from the sampled received signal.
\vspace{-2mm}
\section{Complexity and Numerical Studies}
\vspace{-2mm}
In all subsequent analyses, complexity associated with the multiplication operation is considered only. First, we will analyze complexities of the block implementations of WLE, LE and ROE: where sizes of the referred matrices and vectors are $k\times k$ and $k\times 1$, respectively.  Computational load of WLE depends mainly in computing $\textbf{C}$ from (\ref{Find-C}), where the complex matrix inversion dominates.
If the matrix was real, complexity would be $O(k^3)$ \cite{Intro_to_Algo}. On the contrary, a complex matrix inversion is an equivalent multiplication of two complex matrices \cite{Intro_to_Algo}. Thus, effective computations can be approximated as $O(4k^3)$.
Complexity of LE will be slightly less than that of WLE. However, since there is a complex inversion, as given by \eqref{Filter-LE}, the computational load is still roughly $O(4k^3)$.
Inversion of the real matrix in (\ref{estimate-a}) primarily dominates the complexity of ROE.
Hence, $O(k^3)$ is the approximate complexity of ROE.

\begin{figure}[b]
\centerline{\includegraphics[height=4 cm, width=0.9 \columnwidth]{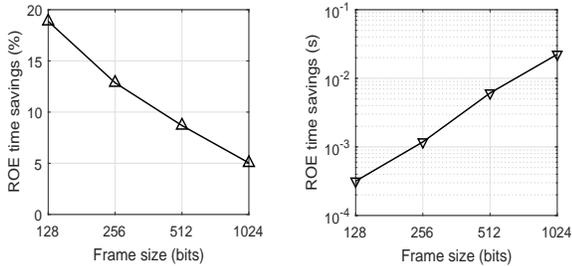}}
\vspace{-2mm}
\caption{Time complexity comparison between ROE and WLE.}
\label{Plot-Complexity}
\vspace{-2mm}
\end{figure}

Looking at the dominating terms, it may be inferred that ROE is expected to be superior to WLE in terms of complexity.
For illustration, we have simulated the relative time complexities in MATLAB, for some practical frame lengths. For purposes of numerical analyses, OQPSK is used as a special case of the OQAM scheme. The workstation used was equipped with a 4th generation core-i7, 2.2 GHz processor. Fig. \ref{Plot-Complexity} depicts the results obtained. The plot at the left shows savings achieved by ROE, when expressed as a percent of WLE execution time. The graph to the right displays actual savings in execution times. As can be observed, ROE provides up to about 20 percent savings in execution time over WLE. This percentage gain decreases with longer frame lengths. However, it is important to note that optimal frame lengths for contemporary environments, such as ZigBee in internet of things (IoT), is less than 1024 bits \cite{Frame-Length}. Finally, even though percentage gain decreases, savings in absolute time increases for ROE at higher frame lengths.

In NLMS implementations, corresponding complexities of ROE, WLE and LE are outlined in Table \ref{tab1}.
It is noticeable that ROE has the least complexity, and it requires less than one-third the number of computations needed for WLE.

\begin{table}[b]
\caption{Complexity comparison in NLMS implementations.}
   \footnotesize
\begin{center}
\begin{tabular}{|c|c|c|c|c|}
\cline{1-5}
To Get: & $\hat{\textbf{x}}(n)$ & $\tilde{\mu}(n)$ & $\textbf{w}_{n+1}$ & Total \\
\hhline{|=|=|=|=|=|}
\text{ROE} & $4mL_b+2$ & $4mL_b+3$ & $4mL_b+3$ & $12mL_b+8$ \\
\hline
\text{WLE} & $16mL_b+8$ & $8mL_b+5$ & $16mL_b+9$ & $40mL_b+22$ \\
\hline
\text{LE} & $8mL_b+8$ & $4mL_b+3$ & $8mL_b+5$ & $20mL_b+16$ \\
\hline
\end{tabular}
\label{tab1}
\end{center}
\end{table}

\begin{figure}[t]
\centerline{\includegraphics[height=5.0 cm, width=\columnwidth]{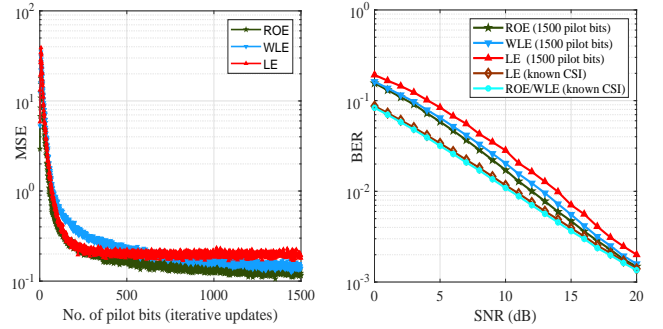}}
\vspace{-2mm}
\caption{Convergence and performance comparisons among the equalizers.}
\label{Plot-LMS}
\vspace{-2mm}
\end{figure}

Results from a numerical study of the NLMS equalizers have been illustrated in Fig. \ref{Plot-LMS}: under randomly generated 3-tap (adjacently located) exponentially decaying channels, 1500 data bits, and $m=3$. The graph to the left illustrates convergence of the adaptive equalizers, by plotting the mean squared error (MSE) for 1500 training bits, at an SNR of 10dB. Rate of convergence and steady-state errors are both functions of the convergence parameters $\mu_\text{LE/ROE/WLE}$ and $\delta_\text{LE/ROE/WLE}$. After an exhaustive empirical search, the numerical value 1 was found optimum for $\mu_\text{LE/ROE/WLE}$, while 2 and 16 were optimal for the $\delta_\text{ROE/WLE}$ and $\delta_\text{LE}$, respectively. This optimality was assessed in terms of finding the minimum steady-state errors for all the equalizers, for the given training sequence.
The graph to the right demonstrates comparative BER performances among all the equalizers. ROE performs better than both LE and WLE in the adaptive domain in terms of BER, owing to its better convergence.
Aside, as expected, BER performances of the NLMS equalizers suffer compared to their block implementations with known CSI.
\vspace{-2mm}
\vspace{-2mm}
\section{Conclusion}
\vspace{-1mm}
In this correspondence, ROE (real-only equalizer) for the OQAM signaling scheme has been introduced.
We have proven that output of the new ROE is equivalent to that of the popular WLE, in a complex channel. Hence, these equalizers exhibit the same performance, and both outperform the LE.
Our complexity analysis finds that ROE is expected to be less complex than WLE.
Using OQPSK as a special case of OQAM, the numerical results demonstrate that significant time savings, for up to 20 percent for practical frame lengths, can be achieved by ROE over WLE.
For a given number of training symbols in the adaptive NLMS implementations (without a matched filter to the complex channel), ROE offers lower BER and also at least 3 times savings in computations, than WLE. Thus, ROE here performs better than WLE, which in turn performs better than LE.
Finally, in the case of a real channel, all the three equalizers exhibit the same performance, with a matched filter to the real channel.
\vspace{-2mm}
\section*{Appendix}
\vspace{-2mm}
Let the output of the WLE be given by
\begin{equation}\label{WLE}
  \hat{\textbf{x}}_\text{WLE} = \textbf{Cy}+\textbf{D} \textbf{y}^*,
\end{equation}
where $\textbf{C}$ and $\textbf{D}$ are WLE coefficients.
\noindent
The orthogonality principle in \cite{picinbono} states that the estimation error is orthogonal to the sampled matched filter vector. From that we obtain
\vspace{-1mm}
\begin{equation}\label{Orth_1}
E[(\textbf{x}-\hat{\textbf{x}})\textbf{y}^{\dagger}] = E[(\textbf{x}-\textbf{C}\textbf{y}-\textbf{D}\textbf{y}^*)(\textbf{x}^{\dagger} \boldsymbol{\Gamma}^{\dagger}+\textbf{v}^{\dagger})]=0,
\end{equation}
which simplifies to
\vspace{-2mm}
\begin{flushleft}
$E[\textbf{x}\textbf{x}^{\dagger}\boldsymbol{\Gamma}]-\textbf{C}E[(\boldsymbol{\Gamma} \textbf{x}+\textbf{v})(\textbf{x}^{\dagger}\boldsymbol{\Gamma}+\textbf{v}^{\dagger})]$
\end{flushleft}
\vspace{-3mm}
\begin{flushright}
$-\textbf{D}E[(\boldsymbol{\Gamma}^* \textbf{x}^*+\textbf{v}^*)(\textbf{x}^{\dagger}\boldsymbol{\Gamma}+\textbf{v}^{\dagger})]= 0,$\\
\end{flushright}
\vspace{-1mm}
where $E[\textbf{x}\textbf{x}^{\dagger}] = \textbf{I}, E[\textbf{x}^* \textbf{x}^{\dagger}] = \textbf{B}, E[\textbf{v}\textbf{v}^{\dagger}] = \sigma^2, E[\textbf{v}^*\textbf{v}^{\dagger}] =0\,.$
Knowing that $\boldsymbol{\Gamma}^* = \boldsymbol{\Gamma}^\top$, (\ref{Orth_1}) becomes
\begin{equation}\label{D-find}
\textbf{D} = [\textbf{I}-\textbf{C}(\boldsymbol{\Gamma} +\sigma^2 \textbf{I})]\textbf{B}[\boldsymbol{\Gamma}^\top]^{-1}.
\end{equation}
\vspace{-1mm}
The orthogonality principle also yields
\begin{equation}\label{Orth_2}
E[(\textbf{x}-\hat{\textbf{x}})^*\textbf{y}^{\dagger}] = E[(\textbf{x}^*-\textbf{C}^*\textbf{y}^*-\textbf{D}^*\textbf{y}) (\textbf{x}^{\dagger}\boldsymbol{\Gamma}^{\dagger}+\textbf{v}^{\dagger})] = 0,
\end{equation}
which simplifies to
\vspace{-2 mm}
\begin{center}
$\textbf{B}\boldsymbol{\Gamma}-\textbf{D}^*(\boldsymbol{\Gamma} +\sigma^2 \textbf{I})\boldsymbol{\Gamma}-\textbf{C}^*[\boldsymbol{\Gamma}^* \textbf{B}] \boldsymbol{\Gamma} = 0.$\\
\end{center}
\vspace{-1mm}
Replacing $\textbf{D}$ and post-multiplying with $(\boldsymbol{\Gamma}^\top)$, we get
\vspace{-1mm}
\begin{equation}\label{Find-C}
\textbf{C} = [\textbf{B} \boldsymbol{\Gamma}^\top \textbf{B} + \boldsymbol{\Gamma} + \sigma^2 \textbf{I}]^{-1}.
\end{equation}
Recall that  (\ref{D-find}) can also be written as
\vspace{-1mm}
\begin{center}
$\textbf{D} = [\textbf{C}\textbf{C}^{-1}-\textbf{C}(\boldsymbol{\Gamma} +\sigma^2 \textbf{I})]\textbf{B}[\boldsymbol{\Gamma}^\top]^{-1},$\\
\end{center}
\vspace{-1mm}
and replacing $\textbf{C}^{-1}$ from (\ref{Find-C}) gives
\begin{equation}
\textbf{D} = \textbf{C}[\textbf{B} \boldsymbol{\Gamma}^\top \textbf{B} + \boldsymbol{\Gamma} + \sigma^2 \textbf{I}-\boldsymbol{\Gamma} -\sigma^2 \textbf{I}]\textbf{B}[\boldsymbol{\Gamma}^\top]^{-1}=\textbf{C} \times \textbf{B}.
\end{equation}
\newpage
\bibliographystyle{IEEEtran}
\bibliography{My_Refs}

\end{document}